\documentclass{article}

\usepackage{enumerate}
\usepackage{amsfonts}
\usepackage{amsmath}
\usepackage{textcomp}
\usepackage{amsthm, tikz}
\usepackage{amssymb}
\usepackage{color}
\usepackage{graphicx}
\usepackage[matrix,arrow,curve]{xy}
\usepackage{array}
\usepackage{mathtools}

\def\be{\begin{eqnarray}}
\def\ee{\end{eqnarray}}
\def\nn{\nonumber}

\def\b{\beta}
\def\m{\mu}

\def\r{\rho}
\def\tr{{\rm tr}\,}
\def\Tr{{\rm Tr}\,}

\def\H{\mathcal{H}}
\def\K{\mathcal{K}}

\def\S{\mathcal{S}}

\def\t{\Delta}

\newdimen\linethick  \linethick=0.4pt
\newdimen\hboxitspace    \hboxitspace=5pt
\newdimen\vboxitspace    \vboxitspace=5pt

\DeclareMathAlphabet{\mathcalligra}{T1}{calligra}{m}{n}

\hoffset= - 1.2in         
\begin{document}
\title{{\bf {On genus expansion of superpolynomials} \vspace{.2cm}}
\author{{\bf Andrei Mironov$^{a,b,c,}$}\footnote{mironov@itep.ru; mironov@lpi.ru}, {\bf Alexei Morozov$^{b,c,}$}\thanks{morozov@itep.ru}, {\bf Alexei Sleptsov$^{b,d,e,}$}\thanks{sleptsov@itep.ru} \ and {\bf Andrey Smirnov$^{b,f,}$}\thanks{asmirnov@math.columbia.edu}}
\date{ }
}

\maketitle

\vspace{-5.0cm}

\begin{center}
\hfill FIAN/TD-14/13\\
\hfill ITEP/TH-41/13\\
\end{center}

\vspace{3.2cm}

\begin{center}
$^a$ {\small {\it Lebedev Physics Institute, Moscow 119991, Russia}}\\
$^b$ {\small {\it ITEP, Moscow 117218, Russia}}\\
$^c$ {\small {\it Moscow Physical Engineering Institute, Moscow 115409, Russia }}\\
$^d$ {\small {\it Laboratory of Quantum Topology,
Chelyabinsk State University, Chelyabinsk 454001, Russia }}\\
$^e$ {\small {\it KdVI, University of Amsterdam, the Netherlands}}\\
$^f$ {\small {\it Columbia University, Department of Mathematics} and
{\it ITEP, Moscow, Russia}} 
\end{center}

\vspace{1cm}

\begin{abstract}
Recently it was shown that the (Ooguri-Vafa) generating function of HOMFLY polynomials is the Hurwitz
partition function,
i.e. that the dependence of the HOMFLY polynomials on representation $R$ is naturally captured by symmetric group characters
(cut-and-join eigenvalues). The genus expansion and expansion through Vassiliev invariants explicitly demonstrate this
phenomenon. In the present letter we claim that the superpolynomials are not functions of such a type: symmetric group
characters do not provide an adequate linear basis for their expansions. Deformation to superpolynomials is, however,
straightforward in the multiplicative basis: the  Casimir operators are $\beta$-deformed to Hamiltonians of the
Calogero-Moser-Sutherland system. Applying this trick to the genus and Vassiliev expansions, we observe that the deformation
is fully straightforward only for the thin knots. Beyond the family of thin knots additional algebraically independent terms appear in the Vassiliev and genus expansions. This can suggest that the superpolynomials do in fact contain more information about knots than the colored HOMFLY and Kauffman polynomials. However, even for the thin knots the beta-deformation is non-innocent: already in the simplest examples it seems inconsistent with {\it the postivity} of colored superpolynomials in non-(anti)symmetric representations, which also happens in I.Cherednik's (DAHA-based) approach to the torus knots.
\end{abstract}

\section{Introduction}

In \cite{MMS1,MMS2} we started a study of the 't Hooft
genus expansion for colored knot polynomials \cite{Kpol}.
There are at least five immediate subjects of interest
about it:
\begin{itemize}
\item
a realization of the AMM/EO topological recursion of \cite{AMMEO},
which was investigated for a couple of simplest examples in \cite{FuDi};
\item
integrability properties of generating functions like the
Ooguri-Vafa partition function \cite{OV}; this includes
their expansion over the symmetric group characters \cite{MMS1,MMS2} and
representation as Hurwitz \cite{MMNhur,MMN}
and ordinary KP/Toda $\tau$-functions \cite{MMM1,MMS2},
relation between these two representations being familiar from the
old theory of renormalization group flows \cite{MMrg};
\item
differential and difference equations \cite{Gar,IMMMfe,MMeqs} satisfied
by the colored knot polynomials as functions of their representation
indices;
\item
relation between the genus and ordinary loop expansions,
where the latter is related to the Vassiliev invariants \cite{VasI,Duzhin} and
the Kontsevich integrals \cite{KoI};
\item
the double-scaling large representation limit, where the knot polynomials
acquire  interpretation in terms of the hyperbolic volumes via
the volume conjecture \cite{vc} and possess interesting ${\cal R}$-matrix
representations \cite{Hik} based on the theory of cluster mutations
\cite{Mu}.
\end{itemize}

Only a partial progress along these directions was achieved in \cite{MMS1,MMS2}
and the entire story requires much more attention than it received so far.
Still, in this letter we proceed in a somewhat {\it different} direction:
we consider the genus expansion of the {\it superpolynomials}
\cite{DGR,sups,sups1,sups2,DMMSS,IMMMfe,AMMM21,Morbetadefo}.
It involves a number of new elements:
\begin{enumerate}[a)]
\item the expansion can no longer be in the symmetric group characters,
and their adequate MacDonald deformation is still unknown;
\item because of this, one can not yet provide a proper modification
of the Hurwitz partition functions; however, the $\beta$-deformation
of the narrower class of Casimir characters is straightforward,
and thus the KP/Toda-like representation is readily available;
\item the loop expansion of superpolynomials can and does (as we demonstrate)
possess more terms than that of HOMFLY, i.e. at $\beta=1$, what can be a signal that
the set of Vassiliev invariants is enlarged by consideration of the
superpolynomials;
\item this observation makes even more important study of the
loop expansion of the Khovanov-Rozansky (KR) polynomials \cite{KhR},
to which the superpolynomials are related in a somewhat non-trivial
fashion\cite{DGR}; this poses new puzzling questions about the
relation between the genus and loop expansions in generic cohomological theories
and also calls for a better understanding of the {\it colored} KR
polynomials, which can hopefully be achieved by unification of modern
alternatives to the cabling method  \cite{AnoMMMpath,AnoM}
and to the KR calculus \cite{DolMor3}.
\end{enumerate}
The goal of the present letter is to begin investigation of these
issues.

\section{Hurwitz exponentials}

\subsection{Hurwitz partition function}

Consider a function $F_R$ given on the space of irreducible representations, which are labeled by the Young diagram $R=\{R_1\geq R_2\geq\dots\geq R_l\}$.
In this section we discuss conditions when this $R$-dependent function $F$ can be expanded into the symmetric group
characters $\varphi_R(\t)$.

Let us introduce variables $m_i = R_i -i$ and assume that $F_R$ is a power series in all variables $\{m_i\}$, and, moreover,
is a symmetric function in $\{m_i\}$. It is well-known that the eigenvalues of Casimir operators for $GL_N$: $C_i$,
$i=1\ldots N$ are symmetric polynomials in $\{m_i\}$ \cite{Zhe}. Moreover, the first $N$ Casimir operators are algebraically
independent and one can easily check that their eigenvalues form a multiplicative basis of symmetric functions of degree not
higher than $N$. Allowing $N$ to be arbitrarily large, one can conclude that $F_R$ can be expanded in the basis of the
eigenvalues of Casimir operators.

Let us make the argument more explicit. Consider an irreducible representation of $GL_N$ labeled by the Young diagram $R$,
then, the Casimir operator is proportional to the identity operator in accordance with the Schur lemma\footnote{The shift $1/2$
can be replaced with any other constant, this induces a linear transformation of the set of Casimir operators, the particular
choice of $1/2$ being more convenient for many purposes, including application to the genus expansion.}:
\be
\hat{C}(k) &=& C_R(k)\hat{I},\\
C_R(k) &=& \sum_{j=1}^{l(R)} (R_j-j+1/2)^k - (-j+1/2)^k.
\label{evC}
\ee
In particular, it follows from the Schur lemma that for the Casimir operator the prime character $\chi_R$ of the linear group is its
eigenfunction:
\be
\hat C(k) \,\chi_R = C_R(k)\chi_R.
\label{caschi}
\ee

It is convenient to introduce multiplicative combinations of Casimir operators $\hat C(\t)$ labeled by partitions (Young
diagrams) $\Delta=\{\delta_1\geq \delta_2\geq\dots\geq \delta_l\}$:
\be
\label{casim}
\hat C(\t) &=& \prod_{j=1}^{l(\t)} \hat C(\delta_i), \\
\hat C(\t) \,\chi_R &=& C_R(\t)\chi_R.
\ee
The symmetric polynomials $C_R(\t)$ of $\{m_i\}$ evidently form a linear basis in the space of all symmetric polynomials of
degree not higher than $N$. Then, each concrete term of the
power series $F_R$ can be expanded in this basis, choosing each time large enough $N$, i.e.
any symmetric power series in $\{m_i=R_i-i\}$ is expanded in the basis of $C_R(\t)$.

Note that, in accordance with the Frobenius theorem,
the linear group character $\chi_R$ is a symmetric polynomial of the monomials $p_{\Delta}$:
\be
p_{\delta} = \sum_ix_i^{\delta}, \ \ p_{\t} = \prod_ip_{\delta_i},\\
\chi_R = \sum_{|\t|=|R|}z^{-1}(\t)\Phi_R^{\t}p_{\t},
\ee
where $z(\t)$ counts the order of the automorphism group of the Young diagram and the transition matrix $\Phi_R^{\t}$ is the
character of the symmetric group \cite{Fulton,Mac}. We use, however, differently normalized
characters $\varphi_R(\t)$ so that
\be\label{charfor}
\chi_R(p) = \sum_{|\t|=|R|}d_R\varphi_R(\t)p_{\t}
\ee
where $d_R$ is dimension of the irreducible representation $R$ of symmetric group \cite{MMN}.

These symmetric group characters $\varphi_R(\t)$ are eigenvalues of the cut-and-join operators $\hat{W}_{\t}$ \cite{MMN}
\be\label{cut}
\hat W_{\Delta}\chi_R = \varphi_{_R}(\Delta)\chi_R.
\ee
and the operators $\hat{W}_{\t}$ form a basis related with that of $\hat C(\t)$ by a linear transformation \cite{MMN}.
Therefore, due to (\ref{casim}) and (\ref{cut}), the eigenvalues of $\hat{C}_{\t}$ and $\hat{W}_{\t}$ are also related by a
linear transformation. Note that the operators $\hat{C}_{\t}$ and $\hat{W}_{\t}$ can be represented by differential operators
in time variables $\{p_k\}$. Their explicit form also can be found in \cite{MMN}.

This claim that the multiplicative combinations of eigenvalues of the Casimir operators (\ref{casim}) are linear
combinations of the symmetric group characters $\varphi_R(\t)$ can be considered as one of the corollaries of
the Schur-Weyl duality.

To summarize aforesaid, one concludes that the function $F_R$ can be expanded in $\varphi_R(\t)$ if and only if $F_R$ is a
symmetric power series in $\{m_i = R_i-i\}$:
\be\label{symm}
\boxed{
{\rm SymmPowSer}\left(R_i-i\right) = {\rm Ser}\{\varphi_R(\t)\}
}
\ee

As is well-known \cite{Dijk} the characters $\varphi_R(\t)$ are related by the Frobenius formula with the Hurwitz numbers
\be
{\rm Cov}_n(\t_1,\dots,\t_k) = \sum_R d_R^2\varphi_R(\t_1)\dots \varphi_R(\t_k)\delta_{|R|,n},
\ee
which counts the number of $n$-sheet coverings of the Riemann sphere with $k$ ramification points of given ramification
types. The type of ramification at the point $i$ characterizes the way in which the sheets are glued, and is
labeled by the Young diagram (integer partition of $n$) $\t_i$ of weight $|\t_i| = n$. The generating function of
the Hurwitz numbers which depends on $K$ infinite sets of variables $p^{(i)}_k$, $i=1\ldots K$
and on an infinite set of variables $\{w_\t\}$ is given by
\be\label{Hpf}
Z(p^{(i)}|w_\t) =\sum_Rd_R^2\prod_{i=1}^K\frac{\chi_R(p^{(i)})}{d_R}\exp\left\{\sum_\t w_\t\varphi_R(\t)\right\}=
 \exp\Big( \sum_{\t}w_{\t}\hat W_\t \Big) Z(p,p^{'},p^{''},\dots|0),
\ee
where $\hat W_\t$ acts on any set of variables $p^{(i)}_k$, i.e. on any of the linear characters in the product.
In the case of $K=2$ this partition function becomes a KP tau-function, see \cite{AMMN}.

Now we return to the function $F_R$. It can be re-expanded to the following form:
\be\label{addb}
F_R=\exp\Big\{\sum_{\t} w_{\t}\varphi_R(\t) \Big\}.
\ee
with $w_\t$ being complicated combinations of the expansion coefficients of $F_R$ in the basis of $\varphi_R(\t)$.
Functions of such a type we call Hurwitz exponential since these are exactly the exponentials entering the
Hurwitz partition functions  (\ref{Hpf}).

\subsection{From Hurwitz to KP partition functions and renormalization group}

Let us comment more on the two bases, multiplicative $C_k$ and additive $\varphi_R(\t)$. Note that the
generic Hurwitz exponential spanned by the additive basis (\ref{addb})
gives rise to the generating function which is {\it not} a KP $\tau$-function \cite{AMMN}. However, if the exponential
is spanned by the linear basis (\ref{evC}),
\be
F_R = \exp\Big\{\sum_{k} t_kC_R(k) \Big\}
\ee
(\ref{Hpf}) at $K=1,2$ is the Toda lattice $\tau$-function \cite{AMMN,O} with respect to times $p^{(1,2)}_k$. Moreover,
it is also the Toda chain $\tau$-function at $K=1$ \cite{AMMN} with respect to times $t_k$.

When dealing with these Hurwitz exponentials, one may keep in mind the following analogy with the renormalization group (RG)
and completeness of basis \cite{MMrg}. Let us consider a quantum field theory partition function
\be
Z(G;\varphi_0;t) = \int_{\mathcal{A};\varphi_0}D\phi\exp\left( \dfrac{1}{2}\phi G\phi + A(t;\phi) \right)
\ee
which depends on: (a) the background fields $\varphi_0$; (b) the coupling constants $t$ and (c) the metric $G$.

The coupling constants parameterize the shape of the action
\be
A(t;\phi) = \sum_{n\in B} t^{(n)}\mathcal{O}_n(\phi),
\ee
where the sum goes over some complete set $B$ of functions $\mathcal{O}_n(\phi)$, not obligatory finite or even discrete.
The space $\mathcal{M}\subset\rm{Fun}(\mathcal{A})$ of actions parameterized by the coupling constants $t^{(n)}$, is
referred to as the moduli space of theories. The actions usually take values in numbers or, more generally, in certain rings,
perhaps, non-commutative. The space $\rm{Fun}(\mathcal{A})$ of all functions of $\phi$ is always a ring, but this needs not
be true about the moduli space $\mathcal{M}$, which could be as small a subset as one likes. However, the interesting notion
of partition function arises only if the completeness requirement is imposed on $\mathcal{M}$. There are two different degrees
of completeness, relevant for discussions of partition functions. In the first case (strong completeness), the functions
$\mathcal{O}_n(\phi)$ form a {\it{linear}} basis in $\rm{Fun}(\mathcal{A})$, then $\mathcal{M}$ is essentially the same as
$\rm{Fun}(\mathcal{A})$ itself. In the second case (weak completeness), the functions $\mathcal{O}_n$ generate
$\rm{Fun}(\mathcal{A})$ as a ring, i.e. an arbitrary function of $\phi$ can be decomposed into a sum of {\it{multiplicative}}
combinations of $\mathcal{O}_n$'s. In the case of strong completeness, the notion of RG is absolutely straightforward, but there
is no clear idea how RG can be formulated in the case of weak completeness (which is more relevant for most modern considerations).

In the strongly complete case, the non-linear (in coupling derivatives) equation, even if occurs, can be always rewritten
as a linear equation. In fact, one can easily make a weakly complete model strongly complete, by adding all the newly emerging
operators to the action $A(t;\phi)$, then, if the product $\mathcal{O}_m\mathcal{O}_n$ is added with the coefficient
$t^{(m,n)}$, one has an identity $\partial^2Z/\partial^{(m)}\partial t^{(n)} = \partial Z/\partial t^{(m,n)}$.

The core of the analogy is that $\varphi_R(\t)$ (linear basis) corresponds to a strongly complete set and $C_R(k)$
(multiplicative basis) corresponds to a weakly complete set of operators. Moreover, as we saw
one can lift the multiplicative basis to the linear basis as the weak completeness lift to the strong completeness,
i.e. via introducing new operators (\ref{casim}).

\subsection{HOMFLY polynomial as Hurwitz exponential \cite{MMS1,MMS2}}
\label{homfly}

In s.2.1 we outlined the necessary and sufficient conditions for $F_R$ to be a Hurwitz exponential.
Here we explain why the HOMFLY polynomials are the Hurwitz exponentials. This fact was realized in \cite{MMS1}
by explicit calculations.

We need to check that the HOMFLY polynomials are symmetric functions in $\{m_i\}$ and power series in $\{m_i\}$. The first
property follows from the fact that the HOMFLY polynomial is the vacuum expectation value of the Wilson loop in $3d$
Chern-Simons theory with the gauge group $SU(N)$ \cite{CS}:
\be
\H_R = \langle\Tr_R(U)\rangle_{_{\rm CS}}=\langle\chi_R(U)\rangle_{_{\rm CS}},
\ee
where $U=P\exp\Big(\oint_C {\bf A}_\mu (x)dx^\mu\Big)$ is the Wilson loop which is a group element of $SU(N)$ and the Chern-Simons
averaging goes over the gauge field ${\bf A}_\mu (x)$. The HOMFLY polynomial $\H_R^\K$ defined in a such way for any knot $\K$ is actually not a polynomial but a rational function. To get a polynomial, one needs to normalize $\H_R$ dividing it by the HOMFLY polynomial of unknot, which is equal to the Schur function $\chi_R(p)$, calculated at the special point (topological locus)
\be
p_k^*={A^k-A^{-k}\over q^k-q^{-k}}.
\ee
Therefore, the normalized HOMFLY polynomial is
\be
H_R^\K = \dfrac{\langle\chi_R(U)\rangle_{_{\rm CS}}}{\chi_R(p^{*})}.
\ee
Using formula (\ref{charfor}) $\chi_R(p) = \sum\limits_{|\t|=|R|}^{}d_R\varphi_R(\t)p_{\t}$ in this case, one gets that
\be
H_R^\K = \dfrac{\Big\langle \sum\limits_{|\t|=|R|}^{}d_R\varphi_R(\t)p_{\t}(U) \Big\rangle_{_{\rm CS}}}{\sum\limits_{|\t|=|R|}^{}d_R\varphi_R(\t)p_{\t}^*} =  \dfrac{ \sum\limits_{|\t|=|R|}^{}\varphi_R(\t) \Big\langle p_{\t}(U) \Big\rangle_{_{\rm CS}}}{\sum\limits_{|\t|=|R|}^{}\varphi_R(\t)p_{\t}^*}
\label{Hchar}
\ee
Since $\varphi_R(\t)$ form a linear basis in the algebra of polynomials symmetric in $m_i=R_i-i$ \cite{Kerov}, it is clear from (\ref{Hchar}) that $H_R^\K$ is a symmetric function in $\{m_i\}$. Moreover, the numerator is a symmetric polynomial and the denominator is a symmetric polynomial. Let us show that the denominator has a constant term as a function of variables $\{m_i\}$. To this end, we rewrite the denominator as
\be
\sum_{|\t|=|R|}\varphi_R(\t)p_{\t}^* = \dfrac{\chi_R(p^*)}{d_R}.
\ee
The dimension $d_R$ is given by the following formula:
\be
d_R = \dfrac{\prod\limits_{1=i<j}^{|R|} \left(m_i-m_j\right)}{\prod\limits_{i=1}^{|R|}\left(m_i+|R|\right)!}
\ee
The character can be expressed through the eigenvalues of matrix $X, \ p_k=\tr X^k$ by the second Weyl determinant formula
\be
\chi_R[X] = \dfrac{\det_{i,j}x_i^{m_j}}{\det_{i,j}x_i^{-j}}.
\ee
From this formula it is obvious that $\chi_R[X]$ has zeros at the points $m_i=m_j, \ \forall i\neq j$, furthermore, the multiplicities of zeros equal to $1$ (take the corresponding derivative). Expanding the function $\det_{i,j}x_i^{m_j}$ as series in $\{m_i\}$ it is easy to prove by induction that the lowest term has exactly the same degree as Vandermond, hence, it is proportional to  $\prod\limits_{1=i<j}^{|R|} \left(m_i-m_j\right)$, which cancels with a similar product in $d_R$. Thus, one would expect that the denominator $\sum\limits_{|\t|=|R|}\varphi_R(\t)p_{\t}^*$ does not have a pole at the point $\{m_i=0\}$ and the HOMFLY polynomial (\ref{Hchar}) is a power series in $\{m_i\}$.

In other words, we claim that the HOMFLY polynomial is a symmetric function of $\{m_i\}$ and can be regarded as a power series in these variables. Therefore, taking into account (\ref{symm}), one can conclude that the HOMFLY polynomial can be expanded into basis of $\varphi_R(\t)$:
\be\label{hurw}
\boxed{
H_R^{\K} = \exp\Big\{ \sum_{\t}w_{\t}^{\K}\cdot\varphi_{_R}(\t) \Big\}
}
\ee
with some coefficients $w_{\t}$.
This means that the $R$-dependence is {\it fully concentrated in} $\varphi_{_R}(\Delta)$, while
the coefficients $w^{\K}_{\t}$ encode the information about the knot $\K$.

Now one can naturally consider the generating function of the {\it non-normalized} HOMFLY polynomials,
i.e. their sums with the Schur
functions depending on the source variables $\bar p_k$:
\be
Z_H=\sum_R\chi_R(\bar p)\H_R=\left< \exp\Big(\sum_n {1\over k}\bar p_k\Tr U^k
\Big)\right>_{_{\rm CS}}=\sum_R\chi_R(\bar p)\chi_R(p^*)H_R
\ee
where the trace is taken over the fundamental representation.
On one hand, this is exactly the Ooguri-Vafa partition function
which was considered in \cite{OV} in the context of duality of the Chern-Simons theory and topological string on
the resolved conifold. On the other hand, this is the Hurwitz partition function (\ref{Hpf}) with $K=2$: the case, when the
Hurwitz partition function celebrates its most interesting properties, in particular, integrability \cite{AMMN}.
Note that the Hurwitz form (\ref{hurw})
of the HOMFLY polynomial is essential not only for studying its
integrable properties \cite{MMS2}, it also provides a link to the open Gromov-Witten invariants etc \cite{Marino}.

In the next two sections we consider two particular expansions of $w^{K}_{\t}$ into perturbative series, which are induced
by two important expansions of the HOMFLY polynomials.

\section{Two important expansions involving Hurwitz structure}

There are several important perturbative expansions of the HOMFLY polynomials. These are: the "volume" \ expansion, genus expansion,
Vassiliev expansion, all of them giving rise to very interesting invariants. We do not consider the "volume" \ expansion here,
because it deals with the limit $|R| \rightarrow \infty$, and we are going to study the $R$-dependence. Hence, we consider
in this section the two other expansions. For each expansion, we specify the coefficients $w^{K}_{\t}$.

\subsection{Genus expansion \cite{MMS1}}

The genus expansion ('t Hooft limit) of HOMFLY polynomials is a perturbative expansion of correlation function commonly
used in matrix model
and QFT, which is also known as large $N$ expansion. It corresponds to coupling constant $\hbar$ of the theory going to zero and
the parameter $N$ of the gauge group $SU(N)$ going to infinity, with their product $\hbar N$ kept constant.
In \cite{MMS1} it was demonstrated that the genus expansion for the HOMFLY polynomials gets natural form of a Hurwitz
exponential (\ref{hurw}) with particular values of constants $w^{K}_{\t}$ naturally scaled with $\hbar^{|\t|+l(\t)-2}$:
\be\label{genhm}
\boxed{
H_R^{\K}(q=e^{\hbar/2},A) = \exp\Big\{\sum_{\t} \hbar^{|\t|+l(\t)-2}\cdot S_{\t}^{\K}(\hbar^2,A)\cdot\varphi_{_R}(\t) \Big\}}
\ee
i.e.
\be
w^{K}_{\t} = \hbar^{|\t|+l(\t)-2}\cdot S_{\t}^{\K}(\hbar^2,A).\hspace{2.35cm}
\ee
The rescaled coefficients $S_{\t}^{\K}(\hbar^2,A)$ are series in $\hbar^2$ with coefficients depending
on the knot $\K$ but not on the representation $R$. Actually, they also depend on the group, e.g. for the Kauffman polynomials
($SO(N)$ group), they are different.

\subsection{Vassiliev expansion}
\label{vasinv}

Though the dependence of HOMFLY polynomials on $R$ is completely captured by $\varphi_R(\t)$, in $S_{\t}^{\K}(\hbar^2,A)$
there is still a dependence on the group, hidden in the parameter $A=q^N$. Therefore, one may ask how to separate
in $S_{\t}^{\K}(\hbar^2,A)$ this
dependence from the dependence on the knot. To answer this question, one has to construct a perturbative expansion of
$S_{\t}^{\K}(\hbar^2,A)$, keeping $N$ finite in (\ref{genhm}), i.e. putting $A=e^{\hbar N/2}$. This leads to the Vassiliev
invariants. In fact, what one obtains in this way is nothing but the Kontsevich integral, which is a perturbative series in
$\hbar$ with coefficients given by the trivalent diagrams and Vassiliev invariants \cite{Labast,DBSS}:
\be\label{vashm}
H_R^{\K}(A=e^{\frac{N\hbar}{2}},q=e^{\frac{\hbar}{2}}) = \sum_{i=0}^{\infty}\hbar^{i}
\sum_{j=1}^{\mathcal{N}_i}  r_{i,j}^{(R)} v_{i,j}^{\K}
\ee
Here $r_{i,j}^{(R)}$ denote the trivalent diagrams, $\mathcal{N}_i$ is dimension of the vector space formed by the trivalent
diagrams, and $v_{i,j}^{\K}$ are the invariants of finite type (Vassiliev invariants) of knot ${\K}$.

Representation (\ref{vashm}) separates the information about the knot and the (gauge) group.
The knot information is given by the Vassiliev invariants which are nothing but the integrals (combinatorial sums) over the
given knot \cite{Labast,DBSS,Viro}. They are rational numbers. The group (and the representation) information is given by
the trivalent diagrams which are generated by the Casimir operators of the corresponding algebra.

The trivalent diagrams are represented by the polynomials of degree $|i|$ in $N$. Thus, (\ref{vashm}) is the double series in
powers of $\hbar$ and $N$ such that the degree of $\hbar$ is larger or equals to that of $N$. As follows from (\ref{genhm}),
the coefficients of these polynomials can be expressed in terms of $\varphi_{_R}(\Delta)$ (see Appendix A for
some examples):
\be\label{homf1}
r_{i,j}^{(R)} &=& \sum_{k=0}^{i} c^{(R)}_{i,j,k}N^k,\nn \\
c^{(R)}_{i,j,k} &=& \sum_{|\Delta|+l(\Delta)-2\leq i-k}c^{\Delta}_{i,j,k} \varphi_{_R}(\Delta),
\label{phidiag}
\ee
where $c^{\Delta}_{i,j,k}\in \mathbb{Q}$.
Representation (\ref{phidiag}) for the trivalent diagrams can be evaluated using paper \cite{MMS1} or by direct calculations.
At the moment, it seems that concrete $c^{\Delta}_{i,j,k}$ have no any particular sense: since there is no a canonical choice
of the basis in the space of trivalent diagrams, one would associate concrete $c^{\Delta}_{i,j,k}$ with a particular choice of
the basis.

Thus, taking into account formulas (\ref{homf1}), one gets
\be\label{vashm2}
\boxed{
H_R^{\K}(\hbar,N) = \sum_{i,j,k}\sum_{|\Delta|+l(\Delta)-2\leq i-k}\hbar^{i}v_{i,j}^{\K}N^kc^{\Delta}_{i,j,k}\varphi_{_R}(\Delta)
}
\ee

\bigskip

\section{Superpolynomial deformation of the Hurwitz exponential}

Superpolynomials are also polynomial invariants depending on the representation $R$, with one additional variable
as compared with the HOMFLY polynomials. Hence, there is a question of describing the $R$-dependence of
superpolynomials, of extrapolating a Hurwitz exponential pattern. In this section, we discuss this question.

\subsection{Superpolynomials are not Hurwitz partition function}

In order to prove that the HOMFLY polynomials are Hurwitz exponentials we used the symmetry property, which follows from the
fact that the HOMFLY polynomials are averages of the characters, that is, of the Schur polynomials. For the superpolynomials,
the Schur polynomials are replaced with the MacDonald polynomials \cite{DMMSS}. However, these latter are not antisymmetric
functions in $\{m_i=R_i-i\}$, i.e. it is impossible to expand the superpolynomial in the basis of $\varphi_R(\t)$.

Consider, for instance, the MacDonald polynomials at the special points $p_k^{*}$. Then, they are manifestly given by
\be
M^*_R = \prod_{k=0}^{\beta-1}\prod_{1\leq j<i\leq N} \frac{q^{R[j]-R[i]}t^{i-j}q^k - q^{R[i]-R[j]}t^{j-i}q^{-k}} {t^{i-j}q^k - t^{j-i}q^{-k}}
\ee
 Put $t=q^\b$. Then, it is clear from this formula that the MacDonald polynomials are antisymmetric in $\{\m=R_i-\b i\}$ rather than $\{m=R_i- i\}$. Moreover, they are uniquely defined as a common system of eigenfunctions of the commuting
set of operators generalizing the Casimir operators, which are nothing but the Ruijsenaars Hamiltonians \cite{Rui}:
\be\label{Ham}
\hat H_{k}=\sum\limits_{i_{1}<...<i_{k}} {1\over\Delta (x)} \hat P_{i_{1}}...\hat P_{i_{k}}
\Delta (x)\, \hat Q_{i_{1}}...\hat Q_{i_{k}},\ \ \ \ [\hat H_{k},\hat H_{m}]=0
\ee
where the Van-der-Monde determinant $\Delta(x)  = \det_{ij} x_i^{N-j} =
 \prod_{i<j}^N(x_i-x_j)$ and the shift operators are defined as:
\be
\hat P_{k}=q^{\beta x_{k} \partial_{x_{k}}}, \ \ \ \hat Q_{k}=q^{(1-\beta) x_{k}\partial_{x_{k}}}
\ee
The spectrum of (\ref{Ham}) can be defined from the eigenvalues of spectral operator:
\be
\left(\sum\limits_{k=0}^{n} z^{k} \hat H_{k} \right)M_{R}(x_{1},...,x_{n})=
\prod\limits_{i=1}^{\infty} (1+z\, q^{R_{i}+\beta(n-i)}) M_{R}(x_{1},...,x_{n})
\ee
Thus, the eigenvalues are symmetric in $\m_i=R_i-\b i$. This appeals to construct a full set of symmetric
polynomials in $\{\m_i=R_i-\b i\}$. Unfortunately, in this case there is no a counterpart of the Schur-Weyl duality known,
and the MacDonald polynomials are not simple characters. Hence, there is no a distinguished set of symmetric functions of
$\m_i=R_i-\b i$, and we construct the full sets of symmetric polynomials in $\{\m_i\}$ in two ways in order to present
the superpolynomial as a deformed Hurwitz exponential.

\subsection{Beta-deformation of Casimir operators}

As a simplest deformation, one may consider a very naive $\b$-deformation of the Casimirs eigenvalues:
\be
C_R^{\b}(k) = \sum_i\left(R_i-\b i+\dfrac{1}{2}\right)^k-\left(\b i+\dfrac{1}{2}\right)^k.
\ee
It is clear that for $\b=1$ they are equal to (\ref{evC}). They are also clearly symmetric in $R_i-\b i$. Hence, one can use them
to construct a full set of symmetric polynomials in analogy with formula (\ref{casim}):
\be
\hat C^{\b}(\t) &=& \prod_{j=1}^{l(\t)} \hat C^{\b}(\delta_i).
\ee
However, any interpretation of this set neither from mathematical nor from physical point of view is known. For this reason, we
suggest another construction.

\subsection{Operators $\hat T_k$}

Since we are interested in a basis symmetric in $\m_i=R_i-\b i$, it is allowed not to depend on $q$ at all. Indeed,
there are "intermediate" symmetric functions, which generalize the Schur functions but still solve the (trigonometric)
Calogero-Moser system. These are the Jack polynomials $J_R$ \cite{Mac}.
They are obtained from the MacDonald polynomials just in the
limit of $q\to 1$. Since the Jack polynomials are defined to be eigenvalues of the
trigonometric Calogero-Moser Hamiltonians $\hat{H}_{i}$,
one can consider their generating function:
\be
\Big( u^{|R|} \, \sum\limits_{i=0}^{\infty} \, u^{-i} \, \hat{H}_{i} \Big) \, J_R={\cal{T}}(u)  J_R
\ee
the generating function for the eigenvalues being
\be
{\cal{T}}(u)=\prod\limits_{(i,j)\in \lambda} \, \Big( u-(i-1) \beta +(j-1)  \Big)
\ee
Then, it is natural to define the operators $\hat T^{\beta}_{k}$ such that
\be
\hat T^{\beta}_{k}J_R &=& T^{\beta}_{k}(R)J_R,\nn\\
T^{\beta}_{k}(R) &=& \sum_{i,j}\Big( (j-1)-\beta(i-1) \Big)^{k-1},
\label{opT}
\ee
These operators play an important role in the theory of Jack polynomials, the latter being somewhat mysteriously connected
\cite{AGT2} with the AGT relations \cite{AGT}.
It is possible to express $C_R^{\b}(k)$ through linear combinations of $T^{\b}_k(R)$.

The full set of functions is then given in complete analogy with (\ref{casim}):
\be\label{opT2}
T^{\beta}_{\t}(R) := \prod_{i=1}^{l(\t)} T^{\beta}_{\delta_i}(R).
\ee
For explicit calculations we used exactly this basis. Thus, one can write the most general form of
the $\beta$-deformation of the Hurwitz exponential (\ref{hurw}) for the superpolynomial:
\be\label{hurwsp}
\boxed{
P_R^{\K} = \exp\Big\{\sum_{\t} \omega_{\t}^{\K}\cdot T^{\beta}_{\t}(R) \Big\}
}
\ee
Particular values of constants $\omega_{\t}$ can be dealt with by particular perturbative
expansions, similarly to the HOMFLY case. We again consider the two particular examples: the
genus expansion and the Vassiliev expansion.

\subsection{Genus expansion of superpolynomials}
The genus expansion of the superpolynomial is given by $\hbar\rightarrow 0$ , $N\rightarrow\infty$, $\hbar N={\rm const}$ (i.e. $A=e^{N\hbar/2}=$const), $\b$ is arbitrary and is a more or less straightforward
generalization of the Hurwitz exponential at $\b=1$, that is,
$\varphi_R(\t)\rightarrow T_R^{\b}(\t)$ and $S(\hbar^2,A)\rightarrow \S(\hbar^2,\b,A)$:
\be\label{gensp}
\boxed{
P_R^{\K}(q=e^{\hbar/2},t=e^{\b\hbar/2},A) \ = \ \exp\Big\{\sum_{\t} \hbar^{|\t|+l(\t)-2}\cdot {\S}_{\t}^{\K}(\hbar^2,\b,A)
\cdot T^{\beta}_{\t}(R) \Big\}}
\ee
i.e.
\be
\omega^{K}_{\t} = \hbar^{|\t|+l(\t)-2}\cdot {\S}_{\t}^{\K}(\hbar^2,\b,A). \hspace{2.35cm}
\ee

Why do we think that this definition makes sense? First of all, making computational experiments with particular superpolynomials, we observe that their expansions are very similar to expansions of the HOMFLY polynomials (\ref{genhm}). By this reason we suggest that the definition of superpolynomial expansion has same structure. Second, we may expect that the dependence on the representation $R$ should be described by nice enough understandable functions. Actually, they should be $\b$-deformed counterparts of symmetric group characters, but at the present moment we are not aware of such a deformation. In the previous subsections, basing on symmetric properties of the Macdonald polynomials, we gave our arguments why we consider just $T^\b_\t(R)$ as a full basis. Third, we expect that in the order $n$ of the expansion only the diagrams satisfing the condition $$|\t|+l(\t) \leq n + 2$$ can contribute. Our definition (\ref{gensp}) is based on these three points.

Originally, there was a hope that the coefficients ${\S}_{\t}^{\K}(\hbar^2,\b,A)$ of the expansion could be independent on $\b$ so that the full $\b$-dependence would be hidden in the group factors $T_\t^\b(R)$. However, it seems not to be the case. Instead, the coefficients ${\S}_{\t}^{\K}(\hbar^2,\b,A)$ are deformations of ${S}_{\t}^{\K}(\hbar^2,A)$ arising in the HOMFLY case, where they are generating functions of the higher special polynomials $${S}_{\t}^{\K}(\hbar^2,A) = \sum\limits_{n=0}^{\infty}\hbar^{2n} \sigma_\t^\K(n).$$ One can proceed similarly in the superpolynomial case (\ref{gensp}) $${\S}_{\t}^{\K}(\hbar^2,\b,A) = \sum\limits_{n=0}^{\infty}\hbar^{2n} s_\t^\K(n)$$ Here $s_\t^\K(n)$ are polynomials in $A$ and $\b$. In Appendix {\bf \ref{apex}}, we give a few explicit examples.

At the zero and the first orders our definition is agreed with the expansion from \cite{AMMM21}:
\be
\label{antsp}
P_R(A,q,t)_{q=e^{\hbar/2},\  t=e^{\beta\hbar/2}} = P_{\Box}^{|R|} + \hbar\left(\nu_{_{R^{t}}}-\b\nu_{_R}\right)\sigma_{\Box}^{|R|-2}\sigma_2 + \ldots,
\ee
where $\sigma_{\Box}$ is the special polynomial, $\sigma_2$ is the higher special polynomial of the first order (see \cite{MMS1} for details), $\nu_{_R}=\sum\limits_iR_i(i-1)$.

Equation (\ref{gensp}) is far from being tautological: it implies non-trivial relations between superpolynomials in different representations. In particular, it restricts the shape of superpolynomial for the hook Young diagrams, as soon as one knows it in all symmetric and antisymmetric representations. As we are going to demonstrate, this restriction is quite severe: it can even imply that generic colored superpolynomials are non-positive! Whether this is a drawback or an advantage of our postulate (\ref{hurwsp}) remains to be understood.

\subsection{Is colored superpolynomial positive?}

There is a lack of examples of superpolynomials in non-symmetric representations. Even so, we check all known examples for consistency with expansion (\ref{gensp}). That is, we expanded the superpolynomials in symmetric and antisymmetric representations into series of $\hbar$ $\left( q=e^{\hbar/2},t=e^{\b\hbar/2} \right)$, and then calculated the coefficients in front of the corresponding $T_k^\b(R)$. For the first and the second orders, it is enough to consider only (anti-)symmetric representations to fix all the coefficients. Thus, the first and the second orders are determined for all other representations. We checked if they coincide with the corresponding orders of known (hypothetical) superpolynomials in non-symmetric representations. We considered the two cases: the trefoil and the figure eight knot. The explicit expressions for the zero, first and second orders are given in Appendix {\bf \ref{apex}}.

In \cite{AMMM21} it was suggested to look for a superpolynomial of the figure eight knot in the form with some coefficient $\alpha$:
\be
P^{4_1}_{[2,1]}(q,t,A) &\stackrel{?}{=}& 1 + \Big({\mathfrak Z}_{3|3}^{(-1|-1)}\!\!+{\mathfrak Z}_{1|1}^{(1|0)}+{\mathfrak Z}_{1|1}^{(0|1)}\Big) + \Big( {\mathfrak Z}_{3|3}^{(0|-1)}{\mathfrak Z}_{1|1}^{(1|0)} + {\mathfrak Z}_{3|3}^{(-1|0)} {\mathfrak Z}_{1|1}^{(0|1)} + \alpha {{\mathfrak Z}_{1|1}^{(1|0)}{\mathfrak Z}_{1|1}^{(0|1)}}\Big) + {\mathfrak Z}_{3|3}^{(0|0)}{{\mathfrak Z}_{1|1}^{(1|0)}{\mathfrak Z}_{1|1}^{(0|1)}} \\
{\mathfrak Z}_{I|J}^{(s|\sigma)}(A) &=& \{Aq^{I+s}t^{-\sigma}\}\{Aq^st^{-J-\sigma}\}, \ \ \ \ \ \{q\} = q - {1\over q}
\ee
It was argued that the positivity restricts this coefficient to be $\alpha = 1 - (q-1/t)(t-1/q) = 1 -  1/4(\beta+1)^2\hbar^2 + O(\hbar^4)$:
\be\label{figsup1}
P^{4_1}_{[2,1]} \stackrel{?}{=} {\bf  {1 \over a^6t^{10}q^{10}} \Big(3a^{10}t^{16}q^{10}+5a^8t^{14}q^{12}+3a^8t^{16}q^{16}+q^{12}t^5a^2+3a^4t^{10}q^{16}+5a^4t^6q^8+ q^{20}a^4t^{12}+ 5q^{12}a^4t^8+ } \nn \\ {\bf + q^{20}a^8t^{18}+2a^4t^8q^{14}+3q^4a^4t^4+2a^6t^{15}q^{20}+2q^2a^6t^6+8q^{14}a^6t^{12}+ a^6t^9q^6+ 8q^{10}a^4t^7 +5q^{14}a^4t^9+} \nn \\ {\bf +q^2a^4t^3+q^{18}a^8t^{17}+q^4a^2t+a^2t^7q^{16}+5q^{14}a^8t^{15}+q^8a^2t^3+5a^6t^7q^4 +8q^8a^6t^9+ a^{12}t^{20}q^{10} + 5q^{16}a^6t^{13}+} \nn \\ {\bf +a^{10}t^{14}q^6+a^6t^{13}q^{14}+a^{10}t^{19}q^{16}+a^2t^6q^{14}+a^4t^2+ q^{10}+a^4t^{11}q^{18}+ 5q^6a^4t^5 +a^8t^8+2a^6t^5+11a^6t^{10}q^{10}+} \nn \\ {\bf +2q^6t^4a^4+2a^8t^{12}q^6+2a^8t^{16}q^{14}+ 3a^8t^{10}q^4+ a^8t^9q^2+ 2a^8t^{12}q^{10}+2q^{18}a^6t^{14}+q^6a^2t^2 +a^6t^7q^6+q^{14}a^2t^5+} \nn \\ {\bf +a^{10}t^{15}q^6+2q^{10}t^3a^2 +q^6a^2t+3a^2t^4q^{10} +a^{10}t^{15}q^8 +a^6t^{11}q^{14} +5a^8t^{12}q^8+2a^6t^9q^{10}+2q^{10}a^4t^6+} \nn \\ {\bf +2a^4t^8q^{10}+2a^8t^{14}q^{10}+5q^6a^8t^{11}+q^{14}a^{10}t^{18}
+a^{10}t^{19}q^{14}+a^{10}t^{13}q^4+a^{10}t^{17}q^{12}+2a^{10}t^{17}q^{10}+} \nn \\ {\bf + 8q^{10}a^8t^{13}+8q^6a^6t^8+8a^6t^{11}q^{12}+2a^6t^{11}q^{10} }\Big)
\ee
However, this is inconsistent with our expansion, which requires that $\alpha = 1- \beta\hbar^2 + O(\hbar^3)$. This behavior could imply that $\alpha = 1 - (q-1/q)(t-1/t)$, which is also symmetric under the permutation $q\leftrightarrow t$ (which is expected for such a highly symmetric knot as $4_1$ in representation $[2,1]$ with symmetric Young diagram). However, such $\alpha$ breaks the positivity: not all the coefficients of the superpolynomial enter with positive integer coefficients\footnote{The positivity is presented in the homological variables ${\bf \{a,q,t\} }$, hence we use them in these examples, while for making expansions we use the Macdonald variables $\{A,q,t\}$. In order to clarify our notation, we list here the superpolynomial of the trefoil in the fundamental representation:
\be
P^{3_1}_{[1]} = {A^2\over t^2}\Big(-{q}^{2}A^{2}+{q}^{2}{t}^{2}+1 \Big)= \bf{ -a^2q^8t\Big(  {q}^{4}{t}^{2}+{q}^{2}{t}^{3}{a}^{2}+1  \Big) }
\ee
}:
\be\label{figsup2}
P^{4_1}_{[2,1]} \stackrel{??}{=} {\bf  {1 \over a^6t^{10}q^{10}}\Big(-a^4t^{10}q^{14}+a^{10}t^{16}q^{10}-q^{10}a^2t^5+a^{10}t^{17}q^{12}+5a^8t^{14}q^{12}+3a^8t^{16}q^{16}-a^4t^6q^6+3a^4t^{10}q^{16}- } \nn \\ {\bf - a^8t^{14}q^{14}+5q^{12}a^4t^8 + q^{20}a^8t^{18}+a^4t^8q^{14}+3q^4a^4t^4+2a^6t^{15}q^{20}+ 8a^6t^{11}q^{12}+2q^2a^6t^6+4q^{14}a^6t^{12} -} \nn \\ {\bf -a^6t^9q^6  +3q^{14}a^4t^9+q^2a^4t^3+3q^6a^8t^{11}+q^{18}a^8t^{17}+ 4q^{10}a^8t^{13}+q^4a^2t+a^2t^7q^{16}+3q^{14}a^8t^{15}+5a^6t^7q^4+} \nn \\ {\bf +8q^8a^6t^9+a^{12}t^{20}q^{10}+ 5q^{16}a^6t^{13}+a^{10}t^{14}q^6-a^6t^{13}q^{14}+ a^{10}t^{17}q^{10}+a^2t^6q^{14}+a^4t^2+q^{10}+a^4t^{11}q^{18}+ } \nn \\{\bf +3q^6a^4t^5+a^8t^8+2a^6t^5+7a^6t^{10}q^{10}+4q^6a^6t^8+q^6t^4a^4+q^{14}a^{10}t^{18}+ a^8t^{12}q^6+a^8t^{16}q^{14}+2q^{18}a^6t^{14}+} \nn \\{\bf +q^6a^2t^2-a^6t^7q^6+q^{14}a^2t^5+q^{10}t^3a^2+a^{10}t^{13}q^4+q^6a^2t+ a^2t^4q^{10}+ a^{10}t^{15}q^8-a^6t^{11}q^{14}-a^{10}t^{15}q^{10}-} \nn \\ {\bf -a^8t^{10}q^6+5a^8t^{12}q^8 +5a^4t^6q^8 +4q^{10}
a^4t^7+a^8t^9q^2+q^{20}a^4t^{12}+q^{12}t^5a^2+ a^{10}t^{19}q^{14} +q^8a^2t^3 +} \nn \\ {\bf + a^{10}t^{19}q^{16}+3a^8t^{10}q^4+a^{10}t^{15}q^6 } \Big)
\ee
This provides an example of the phenomenon which is also observed for the trefoil $3_1$ in representation $[2,1]$: of the two suggested expressions, the positive one from \cite{sups2},
\be\label{guksup}
P^{3_1}_{[2,1]} \stackrel{?}{=} {\bf {1\over q^{10}} \Big(  t^{17}q^{10}a^{12}+(q^{16}t^{14}+q^{10}t^{10}+q^{12}t^{12}+t^8q^6+q^4t^8+t^{11}q^{10}+q^{14}t^{12}+q^8t^{10})t^4a^{10}+ } \\ {\bf +(2t^5q^4+q^6t^6+2t^{11}q^{16}+3t^9q^{12} +t^5q^6+t^7q^{10}+q^{10}t^8+t^{13}q^{20}+t^{10}q^{14}+t^9q^{14}+t^3+3t^7q^8)t^4a^8+ } \nn \\ {\bf+(1+2q^4t^2+t^{10}q^{20}+2q^{16}t^8+ 2q^{12}t^6+ 2t^4q^8+t^4q^{10}+t^5q^{10} +  t^3q^6+t^7q^{14})t^4a^6 }\Big) \nn
\ee
and the DAHA-induced from \cite{sups1}
\be\label{DAHAsup}
P^{3_1}_{[2,1]} \stackrel{??}{=} {\bf {1\over q^{10}} \Big( -q^{10}t^{18}a^{12}-(q^{14}t^{14}+q^{12}t^{12}+q^{10}t^{12}+q^4t^8+q^8t^{10}+q^6t^{10}-q^{10}t^{10} + q^{16}t^{14})t^3a^{10} -} \\ {\bf -(t^9q^{10}-t^7q^{10}+3t^9q^{12} +2t^{11}q^{16}+3t^7q^8-t^9q^{14}+t^{11}q^{14}-t^5q^6+t^{13}q^{20}+2t^5q^4+t^7q^6+t^3)t^3a^8 -} \nn \\ {\bf -(1+2q^4t^2+t^{10}q^{20} +2q^{16}t^8 + t^6q^{10}+2q^{12}t^6-t^6q^{14} +2t^4q^8-q^6t^2-t^4q^{10})t^3a^6 } \Big) \nn
\ee
only the second one is consistent with our expansion (\ref{gensp}). It looks like there is a contradiction between the algebraically constructed superpolynomials in non-(anti)symmetric representation  (a natural name would be
Macdonald superpolynomials) and the hypothetical positive (triply-graded) superpolynomials. This apparent contradiction adds a new intrigue to the story of colored superpolynomials.


\section{Vassiliev expansion for superpolynomials}
Another interesting application of formula (\ref{hurwsp}) is the ordinary loop expansion, in the knot theory also known as Vassiliev expansion. As usual for superpolynomials, there are two essentially different cases: {\it thin} knots and {\it thick} knots \cite{DGR}.

\subsection{Thin knots}

The Vassiliev expansion for the superpolynomial $P_R^{\K}(A|q|t)$ is provided with $\hbar\rightarrow 0$ , $N,\ \beta$ fixed
in the variables:
\be
q=e^{\hbar/2}, \ \ A=e^{N\hbar/2}, \ \  t=e^{\beta\hbar/2}.
\ee
Since when $t=q$ the superpolynomial reduces to the HOMFLY polynomial, one gets (\ref{vashm}) for $\b=1$.

In the case of thin knots this expansion takes the form
\be\label{vassp}
P_R^{\K}(A=e^{\frac{N\hbar}{2}},q=e^{\frac{\hbar}{2}},t=e^{\beta\hbar/2}) =
\sum_{i=0}^{\infty}\hbar^{i} \sum_{j=1}^{\mathcal{N}_{i}^{\beta}}  D_{i,j}^{(R)} v_{i,j}^{\K},
\ee
where $D_{i,j}^{(R)}$ are beta-deformations of trivalent diagrams, $v_{i,j}^{\K}$ {\bf are the same
Vassiliev invariants} as in (\ref{vashm}). Thus, the superpolynomials of the thin knot does not contain any new information
about the knot as compared with the HOMFLY case. However, the structure of group factors is different:
$r_{i,j}\rightarrow D_{i,j}$. To describe it, one needs to construct a beta-deformation
in a suitable way:
\be
D_{i,j}^{(R)} &=& \sum_{k=0}^{i} C^{(R)}_{i,j,k}N^k,\nn \\
C^{(R)}_{i,j,k} &=& \sum_{\t+l(\t)-2\leq i-k}C_{i,j,k}^{(\Delta)} T^{\beta}_{\t}(R).
\label{supdiag}
\ee

Thus, taking into account formulas (\ref{supdiag}), expansion (\ref{vassp}) takes the form
\be\label{vashm3}
\boxed{
P_R^{\K}(\hbar,\beta,N) \ = \ \sum_{i,j,k}\sum_{|\Delta|+l(\Delta)-2\leq i-k}\hbar^{i}v_{i,j}^{\K}N^kC^{\Delta}_{i,j,k}T^{\beta}_{\t}(R)
}
\ee

\subsection{Thick knots}

For the thick knots, the perturbative expansion in $\hbar$ is different from that in the previous subsection:
\be\label{vassp2}
P_R^{\K}(A=e^{\frac{N\hbar}{2}},q=e^{\frac{\hbar}{2}},t=e^{\beta\hbar/2}) =
\sum_{i=0}^{\infty}\hbar^{i} \sum_{j=1}^{\mathcal{N}_{i}^{\beta}}  D_{i,j}^{(R)} v_{i,j}^{\K} \ \ +
\ \ (\b-1)\cdot\sum_{i=0}^{\infty}\hbar^{i} \sum_{j=1}^{\mathcal{M}_{i}^{\beta}}  \Xi_{i,j}^{(R)} \r_{i,j}^{\K},
\ee
where the first sum is the same as for the thin knots, while the second sum is {\bf crucially new}:
$\Xi_{i,j}^{(R)}$ are new group structure factors and $\r_{i,j}^{\K}$ are some numbers different from the Vassiliev invariants,
at first glance. One can ask if $\r_{i,j}^{\K}$ could be also related with the Vassiliev invariants, maybe,
they are some linear combinations of $v_{i,j}^{\K}$. In order to answer this question, we recall a definition of invariants
of the finite type (we follow the text-book \cite{Duzhin}).

A knot invariant is said to be a Vassiliev invariant (or a finite type invariant) of order (or degree) $n$ if its extension
vanishes on all singular knots with more than $n$ double points. A Vassiliev invariant is said to be of order (degree) $n$
if it is of order $n$ but not of order $n-1$. Any knot invariant can be extended to knots with double points by means of the
Vassiliev skein relation:
\begin{figure}[h!]
\centering\leavevmode
\includegraphics[width=7 cm]{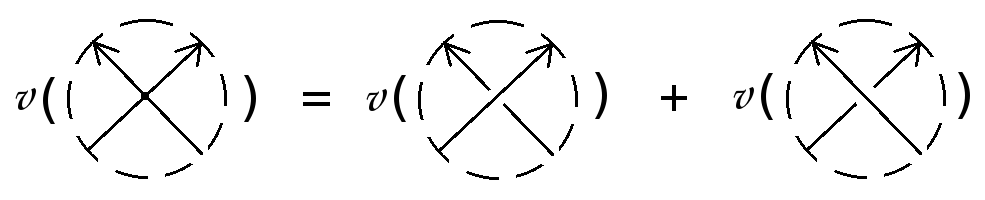}
\caption{Vassiliev skein relation}
\label{vskein}
\end{figure}

Using the Vassiliev skein relation recursively, one can extend any knot invariant to knots with an arbitrary number of double
points. There are many ways to do this, since one can choose to resolve double points in an arbitrary order. However, the result
is independent of the choice.

Applying the skein relation recursively to the simplest thick knot $T[3,4]=8_{19}$ with 4 double points, we have manifestly checked
that $\rho_{2,1}$ and $\rho_{3,1}$ are not the Vassiliev invariants of order 3 at least. It is possible that they are invariants
of higher order, e.g. of 26 or 42. However, this would look quite unusual, since $\rho_{i,j}$ have natural graduation
by powers of $\hbar$ as well as the Vassiliev invariants $v_{i,j}$. This question clearly deserves a further detailed analysis.

\section*{Acknowledgements}

Our work is partly supported by Ministry of Education and Science of the Russian Federation, the Brazil National Counsel of Scientific and Technological Development (A.Mor.), by the program  of UFRN-MCTI (Brazil) (A.Mir.), by NSh-3349.2012.2, by RFBR grants 13-02-00457 (A.Mir. and A.Sl.), 13-02-00478 (A.Mor.), 12-02-00594 (A.Sm.), by joint grants 12-02-92108-Yaf, 13-02-91371-ST, 14-01-93004-Viet, 14-02-92009-NNS and by leading young scientific groups RFBR 12-01-33071 mol-a-ved (A.Sm.). A.Sleptsov is partially supported by Laboratory of Quantum Topology of Chelyabinsk State University (Russian Federation government grant 14.Z50.31.0020).

\newpage

\appendix

\section{Examples of genus expansion for superpolynomials}
\label{apex}

\subsection{Trefoil}
Let us write explicitly a few terms of the genus expansion for the torus knot $T[2,3]$:
\be
P_R(q=e^{\hbar/2},t=e^{\b\hbar/2},A) \ = \ \tilde s_R(1) \cdot \exp\Big\{ \ \hbar\cdot \tilde s_R(2) + \hbar^2\cdot \tilde s_R(3) + \ldots \Big\}
\ee
\be
&\tilde s_R(1) = \left(2-A^2\right)^{|R|} \\ \nn \\
&\tilde s_R(2) = {(A^2-1)(3A^2-5)\over(-2+A^2)^2}T_2^\b(R)  -  {(A^2-1)(\beta-1)\over-2+A^2}T_1^\b(R) \\ \nn \\
&\tilde s_R(3) = -{1\over2}{(8A^2-13)(A^2-1)^2\over(-2+A^2)^4}T_3^\b(R)  +  {(A^2-1)(2A^4-8A^2+7)(\beta-1)\over(-2+A^2)^4}T_2^\b(R) - \nn \\  &-  {1\over2}{(A^2-1)\left((1-\beta+\beta^2)A^4+(-4\beta^2-4+6\beta)A^2+4\beta^2-7\beta+4\right)\over(-2+A^2)^4}T_1^\b(R)  -  {1\over2}{(-9A^4+21A^2+A^6-15)\beta\over(-2+A^2)^4} \left(T_1^\b(R)\right)^2
\ee

The expansion of the DAHA-superpolynomial (\ref{DAHAsup}) agrees with these formulas, while the expansion of superpolynomial (\ref{guksup}) disagrees even at the first order.

\subsection{Figure eight knot}
Let us write explicitly a few terms of genus expansion for the figure eight knot:
\be
P_R(q=e^{\hbar/2},t=e^{\b\hbar/2},A) \ = \ \tilde s_R(1) \cdot \exp\Big\{ \ \hbar\cdot \tilde s_R(2) + \hbar^2\cdot \tilde s_R(3) + \ldots \Big\}
\ee
\be
&\tilde s_R(1) = \left( {A^4-A^2+1\over A^2} \right)^{|R|} \\ \nn \\
&\tilde s_R(2) = {(A^4-1)(2A^4-3A^2+2) \over (A^4-A^2+1)^2}T_2^\b(R) - {1\over2}{(\beta-1)(A^4-1)\over A^4-A^2+1}T_1^\b(R)  \\ \nn \\
&\tilde s_R(3) = -{1\over2}{A^2(A^8-19A^6+29A^4-19A^2+1)(A^2-1)^2 \over (A^4-A^2+1)^4}T_3^\b(R) + {3\over2}{A^2(A^8-2A^6+A^4-2A^2+1)(A^2-1)^2(\beta-1)\over (A^4-A^2+1)^4}T_2^\b(R) -  \\ &-   {1\over2}{(2A^{12}-3A^{10}+4A^6-3A^2+2)\beta A^2 \over (A^4-A^2+1)^4}\left( T_1^\b(R) \right)^2 - {3\over8}{A^2(A^2-1)^2(A^8+\beta^2A^8-2\beta A^8-2\beta^2A^6-2A^6+2\beta A^4+3\beta^2A^4+3A^4-2\beta^2A^2-2A^2+1+\beta^2-2\beta)\over (A^4-A^2+1)^4}T_1^\b(R) \nn
\ee

The expansion of superpolynomial (\ref{figsup2}) agrees with these formulas, while the expansion of superpolynomial (\ref{figsup1}) agrees at the first order and disagrees at the second one.


\section{Beta-deformation of group structure}

In this Appendix we present explicit expressions for a few trivalent diagrams and their beta-deformed generalizations.
As we discussed in Section (\ref{vasinv}), the trivalent diagrams are linear basis in the space of chord diagrams.
Graphically they are represented as follows:
\begin{figure}[h!]
\centering\leavevmode
\includegraphics[width=7 cm]{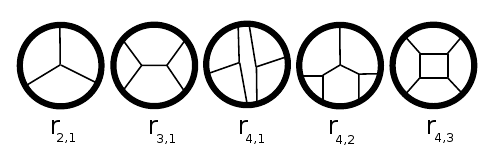}
\caption{Trivalent diagrams}
\label{p_chords}
\end{figure}
and algebraically they are equal to
\be\label{triv}
r_{2,1}^{(R)} &=& \dfrac{1}{4}\left(-|R|{\cdot}N^2 - 2\varphi_{_R}([2]){\cdot}N + |R|^2\right)\\
r_{3,1}^{(R)} &=& \dfrac{1}{8}N\left(|R|{\cdot}N^2 + 2\varphi_{_R}([2]){\cdot}N - |R|^2\right)\\
r_{4,1}^{(R)} &=& \dfrac{1}{16}\left(|R|^2{\cdot}N^4 + 4|R|\varphi_{_R}([2]){\cdot}N^3 + 2(2\varphi_{_R}^2([2])-|R|^3){\cdot}N^2 - 4\varphi_{_R}([2])|R|^2{\cdot}N + |R|^4\right)\\
r_{4,2}^{(R)} &=& \dfrac{1}{16}N^2\left(-|R|{\cdot}N^2 - 2\varphi_{_R}([2]){\cdot}N + |R|^2\right)\\
r_{4,3}^{(R)} &=& \dfrac{1}{16}\Big(|R|{\cdot}N^4 + 6\varphi_{_R}([2]){\cdot}N^3 + 16\left(\frac{3}{4}\varphi_{_R}([3]) + \frac{7}{8}\varphi_{_R}([1,1]) + \frac{1}{16}\varphi_{_R}([1]) \right){\cdot}N^2 -\\&-& 16\left(\frac{1}{2}\varphi_{_R}([4]) + \varphi_{_R}([2,1]) + \frac{1}{2}\varphi_{_R}([2]) \right){\cdot}N -  \Big(2\varphi_R([1])+28\varphi_R([1,1])+72\varphi_R([1,1,1])+24\varphi_R([3,1])-48\varphi_R([2,2])\Big) \Big) \nn
\ee
The beta-deformations of these formulas are
\be
D_{2,1}^{(R)} &=& \dfrac{1}{4}\left( -|R|{\cdot}N^2 - 2\left(T^{\beta}_2-\dfrac{1}{2}(\b-1)T_1 \right){\cdot}N + \beta|R|^2 \right)\\
D_{3,1}^{(R)} &=& \dfrac{1}{4}\left( -2N+1-\b \right)D_{2,1}^{(R)}\\
D_{4,1}^{(R)} &=& \Big( D_{2,1}^{(R)} \Big)^2 \\
D_{4,2}^{(R)} &=& \Big( \dfrac{1}{4}\left( -2N+1-\b \right) \Big)^2 D_{2,1}^{(R)}\\
D_{4,3}^{(R)} &=& \dfrac{1}{16}\Big(|R|{\cdot}N^4 + 6\left(T^{\beta}_2-\dfrac{1}{2}(\b-1)T_1 \right){\cdot}N^3 + 16\left( \frac{3}{4}T^{\beta}_3 - \frac{5}{8}(\beta-1)T^{\beta}_2 + \frac{7}{64}(\beta-1)^2T_1 + \frac{1}{16}\beta T_1^2 \right){\cdot}N^2 -\\&-& 16\left( \frac{1}{2}T^{\beta}_4 - \frac{15}{16}(\beta-1)T^{\beta}_3 + \frac{7}{32}(\beta-1)^2T^{\beta}_2 + \frac{1}{16}(\beta-1)\left(T^{\beta}_2\right)^2 - \frac{1}{64}(\beta-1)(3\beta^2-2\beta+3)T_1  \right){\cdot}N - ?   \Big) \nn
\ee

\section{Examples of new invariants $\rho_{i,j}$}

Here we list the Vassiliev invariants and new invariants for a few series of torus knots.
\be
\begin{array}{|c|c|c|}
\hline
&&\\
\K&v_{3,1}&\rho_{3,1}\\
&&\\
\hline
&&\\
T[3,3k+1]&4k(3k+1)(3k+2)&-\dfrac{k^2(k+1)}{4}\\
&&\\
\hline
&&\\
T[3,3k+2]&4(k+1)(3k+1)(3k+2)&-\dfrac{k(k+1)^2}{4}\\
&&\\
\hline
&&\\
T[4,4k+1]&\dfrac{80k(2k+1)(4k+1)}{3}&-\dfrac{k^2(4k+3)}{2}\\
&&\\
\hline
&&\\
T[4,4k+3]&\dfrac{80(k+1)(2k+1)(4k+3)}{3}&-\dfrac{(k+1)^2(4k+1)}{2}\\
&&\\
\hline
&&\\
T[5,5k+1]&\dfrac{100k(5k+1)(5k+2)}{3}&-\dfrac{7k^2(5k+3)}{4}\\
&&\\
\hline
&&\\
T[5,5k+2]&\dfrac{20(5k+1)(5k+2)(5k+3)}{3}&-\dfrac{k(35k^2+42k+11)}{4}\\
&&\\
\hline
&&\\
T[5,5k+3]&\dfrac{20(5k+2)(5k+3)(5k+4)}{3}&-\dfrac{(k+1)(35k^2+28k+4)}{4}\\
&&\\
\hline
&&\\
T[5,5k+4]&\dfrac{100(k+1)(5k+3)(5k+4)}{3}&-\dfrac{7(k+1)^2(5k+2)}{4}\\
&&\\
\hline
\end{array}
\ee

\end{document}